\documentstyle[12pt]{article}
\begin{document}
\title{ HBT correlations and charge ratios in multiple production of pions}
\author{A.Bialas and K.Zalewski\thanks{Also at the Nuclear Physics Institute,
Krakow, Poland}
\\ M.Smoluchowski Institute of Physics
\\ Jagellonian University, Cracow\thanks{Address: Reymonta 4, 30-059 Krakow,
Poland; e-mail: bialas@thp4.if.uj.edu.pl, zalewski@chall.ifj.edu.pl}}
\maketitle
\begin{abstract}
The influence of the HBT effect on the multiplicity distribution and charge
ratios of independently produced pions is studied. It is shown that, for a
wide class of models, there is a critical point, where the average number of
pions becomes very large and the multiplicity distibution becomes very broad.
In this regime unusual charge ratios ("centauros", "anticentauros") are
strongly enhanced. The prospects for reaching this regime are discussed.
\end{abstract}

\section{Introduction}

It is now well established that the HBT correlations \cite{ha1} influence
significantly
the momentum distribution of particles created in high-energy collisions. The
effect on
multiplicity
distributions and on particle ratios, however, although predicted theoretically
\cite{pr1}-- \cite{wi1},
has not yet been found in accelerator experiments \cite{bj1}.
In view of the increasing interest in measurements of multiplicity
distributions
in  future collider experiments, we found it worthwhile to reexamine this
problem once more.

We begin by reviewing the basic ideas of the HBT effect, as applied to
processes
of particle production. This will permit to explain our assumptions and to
introduce the notation\footnote{We will follow the approach of \cite{bi1}.}.

   Let
 $\psi_0 (q_1,q_2,...q_n,\alpha)  \equiv \psi_0(q,\alpha)$ be the probability
amplitude for the production of n particles with momenta $[q_1,q_2,...,q_n]
\equiv q$ {\it calculated ignoring the identity of particles}. Here
$\alpha$ denotes a collection of all other quantum numbers which may be
relevant to the process in question ( e.g., the momenta of other
particles which we do not wish to consider explicitly in a "semi-inclusive"
measurement). The density matrix
\begin{equation}
\rho_0(q,q')= \int d\alpha \psi_0(q,\alpha) \psi_0^*(q',\alpha)   \label{1}
\end{equation}
 gives all the available information about the system in question.
In particular, the momentum  spectrum of particles is
\begin{equation}
\Omega_0(q)= \int d\alpha \mid \psi_0(q,\alpha) \mid ^2  = \rho_0(q,q).
\label{2}
\end{equation}
In the following we shall assume that $\Omega_0(q)$ is normalized to 1:
\begin{equation}
\int dq \Omega_0(q)=1 .  \label{2a}
\end{equation}
Suppose now that the particles are identical. In this case the states with
different permutation of particle momenta are non-distinguishable and the wave
function is a sum over all permutations
\begin{equation}
 \psi (q,\alpha) =\sum_P \psi_0 (q_P,\alpha),      \label{3}
\end{equation}
where $q_P$ is the set of momenta $[q_1,q_2,...,q_n]$ ordered according to the
permutation $P$ of $[1,2,...,n]$. Using (\ref{1})-(\ref{3}) we obtain
for the distribution of momenta of identical particles
\begin{equation}
\Omega(q)= \frac1{n!} \int d\alpha \mid \psi(q,\alpha) \mid ^2
 =\frac1{n!} \sum_{P,P'} \rho_0(q_P,q_{P'}).
\label{4}
\end{equation}
The factor $\frac1{n!}$ takes care of the fact that the phase-space for $n$
identical particles is $n!$ times smaller than the phase-space of
the non-identical ones.

Eq.(\ref{4}) summarizes the effect of the identity of particles on the observed
spectra. It is seen that to evaluate this effect it is not enough to
know the spectrum $\Omega_0(q)$. The  full density
matrix $\rho_0(q,q')$ is necessary. Inverting this statement we observe
that  the measurements of the spectra of identical particles provide
information on the density matrix $\rho_0(q,q')$, inaccessible otherwise.
This is, in fact, the reason why these measurements are  are so attractive.

There are three points   worth observing in (\ref{4}).

(i) When momenta of all particles are equal, we obtain the simple
result
\begin{equation}
\Omega(q_1=q_2=...=q_n)= n! \Omega_0(q_1=q_2=...=q_n).
\label{4a}
\end{equation}

(ii) The normalization of $\Omega(q)$ is different from that of $\Omega_0(q)$.
Generally we have
\begin{equation}
\int dq \Omega(q) \geq \int dq \Omega_0(q)
\label{4b}
\end{equation}
Since the relation between $\Omega(q)$ and $\Omega_0(q)$ depends
on $n$, the HBT effect affects not only the shape of momentum
spectrum
but also the multiplicity distribution. This aspect of the problem is the
subject of the present investigation.

(iii) If particles are emitted in a pure state, i.e., if $\rho_0(q,q')=
\psi_0(q)\psi^*_0(q')$ and if $\psi_0(q)$ is symmetric with respect to
interchange of any pair of momenta, we have
\begin{equation}
\Omega(q)= n! \Omega_0(q)
\label{4c}
\end{equation}
a really dramatic result.\footnote{We emphasize again that $\Omega_0(q)$ is the
spectrum calculated with {\it identity of particles being ignored}. Some
authors \cite{we1}, while discussing the HBT effect for pure states, include
the identity
of particles already in calculation of $\Omega_0(q)$. In this case one obtains
of course $\Omega_(q)=\Omega_0(q)$.}

\section{HBT phenomenon in uncorrelated emission}

Consider now a system of $n$ particles emitted independently.
 If we ignore the identity of particles, independent emission implies that
 the density matrix factorizes
\begin{equation}
\rho_0(q,q')= \prod_{i=1}^n \rho_0(q_i,q_i').   \label{7}
\end{equation}
Introducing this into (\ref{4}) we have
\begin{equation}
\Omega(q)
 =\frac1{n!} \sum_{P,P'}\prod_{i=1}^n \rho_0((q_P)_i,(q_{P'})_i).
\label{8}
\end{equation}
Here we are interested in the integral
\begin{equation}
W_n = \int dq \Omega(q)=
\sum_{P}\int \prod_{i=1}^n \left(d^3q_i\rho_0(q_i,(q_{P})_i)\right).
\label{9}
\end{equation}
To calculate $W_n$ we observe that, for each permutation $P$ the integral on
the
right hand side of (\ref{9}) factorizes into a product of contributions from
all the cycles of P (as is well-known, each permutation can be decomposed
into cycles). Let us denote the contribution from a cycle of length $k$ by
$C_k$. We have
\begin{equation}
C_k= \int d^3q_1 ... d^3q_k \rho_0(q_1,q_2) \rho_0(q_2,q_3)
...\rho_0(q_{k-1},q_k)\rho_0(q_k,q_1)   \label{10}
\end{equation}
It follows from (\ref{2a}) that $C_1=1$. For $k>1$, $C_k$
 depends on the form of $\rho_0(q,q')$ and cannot be calculated without further
assumptions. One can prove, however, that all $C_k$ are positive.
Indeed, one
sees from (\ref{10}) that
\begin{equation}
 C_k=Tr[\rho_0]^k.   \label{10a}
\end{equation}
 Since $\rho_0(q,q')$, being a
density matrix, has only non-negative eigenvalues and trace one, $C_k>0$.

 The rest of the calculation is just combinatorics.

We observe first that any two permutations which have identical partitions into
cycles give equal contributions. Denoting by $n_k$ the number of
occurences of a cycle of lenght $k$ in the  set of permutations considered, the
contribution from all of them can be written as
\begin{equation}
W'_n= \prod_{k=1}^n (C_k)^{n_k} \frac{n!}{(k!)^{n_k}}[(k-1)!]^{n_k}\frac1{n_k!}
= n!\prod_{k=1}^n\frac{\left(\frac{C_k}{k}\right)^{n_k}}{n_k!}.  \label{11}
\end{equation}
In the first equality the first factor is the integral, the second is the
number of partitions of the $n$ particles among the cycles, the third is
the number of ways a cycle can be constructed from $k$ particles and the last
one corrects for the permutations of whole cycles.

$W_n$ is obtained by summing $W_n'$ only over permutations which have
partitions into cycles different from each other. This still leaves a large
number of terms but -for large $n$- this number is much smaller than the
original $n!$.

Until now we have considered a fixed multiplicity. If the multiplicity
distribution calculated with identity of particles ignored is given by
$P_0(n)$, the correct multiplicity distribution of identical particles is
\begin{equation}
P(n) =\frac{ P_0(n) W_n} {\sum_m P_0(m)W_m}   \label{12}
\end{equation}
For independent emission $P_0(n)$ is given by the Poisson distribution
\begin{equation}
P_0(n) = e^{-\nu} \frac{\nu^n}{n!}    \label{13}
\end{equation}
 and we
obtain an elegant formula for the generating function of the multiplicity
distribution:
\begin{equation}
\Phi(z) \equiv \sum_n P(n) z^n = \exp\left(\sum_{k=1}^{\infty} \frac
{\nu^k(z^k-1)C_k}{k}\right).   \label{14}
\end{equation}

\section{General discussion of multiplicity distributions obtained from
independent emission}

In this section we  discuss the general properties of the multiplicity
distributions obtained from the Eq.(\ref{14}).

First, we observe that using well-known properties of the generating
function we obtain from (\ref{14}) for the average
multiplicity
\begin{eqnarray}
<n>= \sum_{k=1}^{\infty} \nu^k C_k,  \label{15}
\end{eqnarray}
and for the correlation coefficients (cumulants)
\begin{equation}
K_p = \sum_{k=p}^{\infty} \frac{(k-1)!}{(k-p)!}  C_k \nu^k. \label{16}
\end{equation}
All the cumulants are positive, because all the $C_k$ are.
 This means in particular that the distribution is always broader than the
Poisson one.

Specific properties of the distributions defined by (\ref{14}) depend, of
course, on the value of $\nu$ and of the cycle integrals $C_k$.

The first  important example we would like to  consider is
when  particles are emitted in a pure state, i.e.
\begin{equation}
\rho_0(q,q')= \psi(q) \psi^*(q').  \label{17}
\end{equation}
It follows from (\ref{10}) that $C_k=1$ for all $k$ and  the
generating function becomes
\begin{equation}
\Phi(z)  = \exp\left(\sum_{k=1}^{\infty} \frac
{\nu^k(z^k-1)}{k}\right) = \frac{1-\nu}{1-\nu z}   \label{18}
\end{equation}
corresponding to the geometric distribution
\begin{equation}
P(n)  = (1-\nu)\nu^n ,\;\;\;\; <n>=\frac{\nu}{1-\nu}            \label{19}
\end{equation}
which, at the critical point $\nu \rightarrow 1$, exhibits the phenomenon of
Einstein
condensation. One sees from this example that the resulting multiplicity
distribution has little to do with the original Poisson one, and
the observed average multiplicity may dramatically differ from the
initial $\nu$. As we shall see, this is a general phenomenon.

Further discussion depends on the assumed shape of the single-particle density
matrix.

The evaluation of $C_k$ is greatly simplified
if one works in the basis where the density matrix $\rho_0(q,q')$ is diagonal.

Let us first discuss the case of a discrete eigenvalue  spectrum. We have

\begin{equation}
C_k=\sum_m  \lambda_m^k , \;\;\;\; C_1=\sum_m  \lambda_m =1.   \label{19g}
\end{equation}
and thus the generating function of the multiplicity distribution can be
represented as a product
\begin{equation}
\Phi(z) = \prod_m \Phi_m(z)    \label{19h}
\end{equation}
where
\begin{equation}
\Phi_m(z) =
\frac{1-\lambda_m\nu}{1-z\lambda_m\nu}
\label{19i}
\end{equation}
are the generating functions of the geometric distribution (if the
eigenvalue $\lambda_m$ is degenerate, the corresponding factor in  (\ref{19h})
appears
$g_m$ times, where $g_m$ is the degeneration factor).

Thus we obtain the average multiplicity
\begin{equation}
<n>= \sum_m \frac{\lambda_m\nu}{1-\lambda_m\nu} \geq
\frac{\lambda_0\nu}{1-\lambda_0\nu}   \label{19iiii}
\end{equation}
and the cumulants
\begin{equation}
K_p = (p-1)!\sum_m\left(\frac{\lambda_m\nu}{1-\lambda_m\nu}\right)^p
\geq (p-1)!\left(\frac{\lambda_0\nu}{1-\lambda_0\nu}\right)^p  \label{19j}
\end{equation}
where
 $\lambda_0$ is  the largest eigenvalue of $\rho_0(q,q')$. One sees
 that
 $\nu\lambda_0 = 1$ is the critical point
of the multiplicity distribution (c.f. \cite{pr1}).

Let us now discuss the case when the eigenvalue spectrum is continuous.
Denoting by $\lambda$  the eigenvalues of $\rho_0(q,q')$ and by
$\sigma(\lambda)$ the spectral function
we have
\begin{equation}
C_k=\int_0^1 \sigma(\lambda) \lambda^k d\lambda.    \label{19a}
\end{equation}
The normalization condition  (\ref{2a}) implies that
\begin{equation}
C_1=\int_0^1 \sigma(\lambda) \lambda d\lambda =1.   \label{19b}
\end{equation}

One sees that the problem reduces to a discussion of a single non-negative
function $\sigma(\lambda)$ defined in the interval $[0,1]$ and satisfying the
condition (\ref{19b}). We also note that the eigenvalues equal to zero do not
contribute to $C_k$. One can thus always add to $\sigma(\lambda)$ a term of the
form $a \delta(\lambda)$ (with an arbitrary positive constant $a$) without
changing the results.

To be more specific, we consider the generic spectral function in the form
\begin{equation}
\sigma(\lambda) = \lambda_0^{-2}
\frac{\Gamma(a+b+3)}{\Gamma(a+2)\Gamma(b+1)}
\left(\frac{\lambda}{\lambda_0}\right)^a(1-\frac{\lambda}{\lambda_0})^b
\label{19gg}
\end{equation}
for $0\leq \lambda \leq \lambda_0$ and zero otherwise.

We obtain
\begin{equation}
C_k= \lambda_0^{k-1}
\frac{\Gamma(a+b+3)\Gamma(a+k+1)}{\Gamma(a+2)\Gamma(a+b+k+2)} \label{19hh}
\end{equation}
and the cumulants
\begin{eqnarray}
K_p= \frac{\Gamma(a+b+3)\Gamma(a+p+1)\Gamma(p) }{\Gamma(a+b+p+2)\Gamma(a+2)}
\nu^p \lambda_{0}^{p-1} F(p,a+p+1;a+b+p+2;\lambda_0\nu) .      \label{19ii}
\end{eqnarray}
  The hypergeometric function $F$ becomes singular at $\nu\lambda_0=1$ when $p$
exceeds $b+1$.

Three special cases are of interest:

(i) $b\rightarrow -1$. In this case $\sigma(\lambda)\rightarrow
\frac1{\lambda_0} \delta(\lambda-\lambda_0)$ and we have
$C_k=\lambda_0^{k-1}$. Thus
\begin{equation}
\Phi(z)= \left(\frac{1-\lambda_0 \nu} {1-\lambda_0
\nu z}\right)^{\frac1{\lambda_0}}
  \label{19e}
\end{equation}
and we recognize the negative binomial distribution with the average $<n>=
\frac{\nu}{1-\lambda_0\nu}$.  For  $\lambda_0\rightarrow 0$ we recover
the Poisson distribution (\ref{13}). As discussed in the previous section,
$\lambda_0\rightarrow 1$ corresponds to the pure state $C_k=1$.

(ii) $a=b=0$. Now we have $C_k= \frac2{k+1}\lambda_0^{k-1}$ and
one finds the following formula for the cumulants of the distribution
\begin{equation}
K_p =2\frac{(p-1)!}{p+1}  \nu^p
\left(\frac{ \lambda_0}{1-\lambda_0\nu}\right)^{p-1}
F(1,2;p+2;\lambda_0\nu). \label{19.f}
\end{equation}
From the know properties of the hypergeometric function \cite{ab1} we deduce
that at the critical point $\lambda_0\nu \rightarrow 1$, $<n>$ diverges
logarithmically, whereas all other cumulants diverge as negative powers of
$(1-\lambda_0\nu)$.

(iii) $b>0$. In this case
 the average multiplicity approaches $0$
at the critical point $\lambda_0\nu \rightarrow 1$.
 The divergences
appear only in  cumulants of order $p>b+1$. Such a distribution is
characterized by a
large probability of no particles at all, accompanied by a small probablity of
production of very many particles. Schematically
\begin{equation}
P(0) \approx 1-\frac{a}{N^{b+1}} ,\;\;\;\; P(n)= \frac{a}{N^{b+1}}
p(n)\;\;\;\;n>0. \label{19ff}
\end{equation}
Here  $N\rightarrow \infty$ is the average  of the distribution $p(n)$.

 To summarize,  the distribution becomes always singular when $\nu\lambda_0$
(i.e.
 the
product of the initial average multiplicity  and the maximal eigenvalue of the
density matrix) approaches $1$.
The character of the singularity at this critical point, however, depends
crucially on the behaviour of the spectral function in the vicinity of
$\lambda_0$.

\section{Gaussian density matrix}

In this section we consider the single particle density matrix of the
gaussian form, discussed already in several papers by Pratt
\cite{pr1,pr2,pr3} (see also \cite{wi1})
\begin{equation}
\rho_0(q,q') = \rho_x(q_x,q_x') \rho_y(q_y,q_y') \rho_z(q_z,q_z') \label{20}
\end{equation}
with
\begin{equation}
 \rho_x(q_x,q_x') = \left(\frac1{2\pi \Delta_x^2}\right)^{\frac12}
e^{-\frac{(q_x^+)^2}{2
\Delta_x^2} -\frac12R_x^2(q_x^-)^2},   \label{21}
\end{equation}
where
\begin{equation}
q^+ \equiv \frac12(q+q'); \;\; q^- \equiv q-q'.
\label{21a}
\end{equation}
 Analogous formulae define $\rho_y$ and $\rho_z$.
As easily seen, $\Delta_x^2$ is the average
value of the square of the $x$-component of the particle momentum, and $R^2_x$
is
the average value of the square of the $x$-coordinate of the particle emission
point. The uncertainty principle implies that for  $i=x,y,z$,
\begin{equation}
R_i \Delta_i \geq \frac12   \label{21*}
\end{equation}

In order to determine the multiplicity distribution, we first find the
eigenvalues of the density matrix (\ref{21}).
  To this end we observe that the eigenfunctions of (\ref{21})
are of the form\footnote{This was pointed out to us by A.Staruszkiewicz.}
\begin{equation}
f_m(q)= e^{-\frac12 \frac{R}{\Delta}q^2} H_m(\sqrt{\frac{R}{\Delta}}q),
\label{32a}
\end{equation}
where $H_m(q)$ is the Hermite polynomial of order $m$. Using (\ref{32a}) it is
not difficult to find the eigenvalues:
\begin{equation}
\lambda_m = \lambda_{rst} = \lambda_0(1- \lambda_x)^r(1- \lambda_y)^s(1-
\lambda_z)^t,\;\;\;\; r,s,t = 0,1,..., \label{32b}
\end{equation}
where $\lambda_0 = \lambda_x\lambda_y\lambda_z$ and for $i = x,y,z$
\begin{equation}
\lambda_i = \frac2{(1+2\Delta_iR_i)}
\label{25}
\end{equation}
Note that (\ref{21*}) implies that $\lambda_0 \leq 1$, as necessary.

The generating function is given by (\ref{19h}),(\ref{19i}) and the cumulants
by (\ref{19j}) with $\lambda_m$ given by (\ref{32b}).

Using (\ref{32b}) and  following the arguments of the previous section
we also  obtain the elegant formula
\begin{equation}
C_k= \sum_m \lambda_m^k = \frac{\lambda_0^k}{
[1-(1- \lambda_x)^k][(1-1- \lambda_y)^k][1-(1- \lambda_z)^k]}.
\label{32c}
\end{equation}
Together with (\ref{16}), this gives for the cumulants
\begin{equation}
K_p=\sum_{k=p}^{\infty} \frac{(k-1)!}{(k-p)!}
\frac{(\nu\lambda_0)^k}{
[1-(1- \lambda_x)^k][1-(1- \lambda_y)^k][(1-\lambda_z)^k]},
\label{32d}
\end{equation}
which diverges, as expected,  at $\nu\lambda_0\rightarrow 1$.

\section{Charge ratios}

It was first pointed out by Pratt \cite{pr1,pr2,pr3} that studies of the charge
ratios may be an effective way to uncover the effects of HBT correlations in
multiparticle systems. This issue can be readily treated  using the
methods developped in Sectons 3 and 4.

 We would like to discuss  independent production of
positive, negative and neutral pions.
The main difficulty in formulating the problem is how to implement
the constraint of charge conservation, as this clearly depends on  the dynamics
of the
production process (a thorough discussion can be found in \cite{pr3}). Here we
restrict ourselves to two  cases which, we believe, illustrate well the main
point: the charge ratios obtained may drastically differ from those expected
from the "uncorrected" distributions.

In the first case the constraint of charge conservation is ignored altogether.
This may be justified if the system of particles we consider is a small part of
a very large system. For the generating function of multiplicity distribution
we thus obtain simply
\begin{equation}
\Phi(z_+,z_-,z_0) = \Phi(z_+)\Phi(z_-)\Phi(z_0)   ,       \label{41}
\end{equation}
where $\Phi(z)$ is is the generating function of the multiplicity distribution
of one of the species. Introducing $n_c = n_+ + n_-$ we have
\begin{equation}
\Phi(z_c,z_0) = \Phi^2(z_c) \Phi(z_0)  ,   \label{42}
\end{equation}
From this equation one can obtain the full joint distribution of charged and
neutral pions by the usual methods. For illustration we just quote the results
for two extreme cases $n_0=0$ ("centauros") and $n_c=0$ ("anticentauros"):
\begin{equation}
P(n_0=0) = \Phi(0)   ,\;\;\;\; P(n_c=0) = [\Phi(0)]^2.     \label{42b}
\end{equation}
Using now the results of the previous section we have
\begin{equation}
P(n_0=0) = \prod_m(1-\lambda_m\nu)    ,\;\;\;\; P(n_c=0) =[P(n_0=0)]^2 .
\label{42c}
\end{equation}
One immediate consequence is that the production of "centauros" must be
larger than that of "anticentauros".
 For the Gaussian density matrix we obtain, according to (\ref{32b}),
\begin{equation}
P(n_0=0) = \prod_{rst}\left[1-\lambda_0\nu(1- \lambda_x)^r(1-
\lambda_y)^s(1-\lambda_z)^t\right].
\label{43}
\end{equation}

The second case we  consider is when the charged particles are produced
in pairs, i.e., the distribution is  of the form
\begin{equation}
P(n_c,n_0) = \frac{[P(n_c/2)]^2 P(n_0)}{\sum_m[P(m)]^2}. \label{44}
\end{equation}
Thus for production of "centauros" we obtain the same formula as before.
The generating function reads
\begin{equation}
\Phi(z_c,z_0) = \Psi(z_c) \Phi(z_0)     \label{44a}
\end{equation}
where
\begin{equation}
\Psi(z_c)=\frac{ \sum_{n_c} [P(n_c/2)]^2z_c^{n_c} }{\sum_m[P(m)]^2}. \label{45}
\end{equation}
The probabilty of creating an "anticentuaro" is
\begin{equation}
\Psi(z_c=0)=\frac{ [\Phi(0)]^2 }{\sum_m[P(m)]^2} >  [\Phi(0)]^2
\label{46}
\end{equation}
and thus the production of "anticentauros" is enhanced as compared to the
previous case.

These results are illustrated in Fig.1, where the probability of the occurence
of a
"centauro", $P(n_0=0)=\Phi(0)$, given by (\ref{43}), is plotted versus the
average
multiplicity of the system considered for different values of the parameter
$R\Delta$,
which determines the maximal eigenvalue $\lambda_0$ through the relation
(\ref{25}). One sees that for $R\Delta \leq 1$ this probability remains
substantial even for rather large values of the total multiplicity. With
increasing $R\Delta$, however, it drops pretty fast even at moderate values of
the average multiplicity.

\section{ Discussion and outlook}

One sees from the results of the previous sections that the effects of quantum
interference can modify substantially the multiplicity distributions expected
naively from an uncorrelated emission of "distinguishable" particles. The
modifications become  spectacular when the system of particles approaches
the critical point: the multiplicity distribution becomes very broad and does
not resemble in any way the original Poisson distribution characteristic of
uncorrelated emission. This means in particular that one expects  relatively
large probabilities for unusual configurations such as "centauro" or
"anticentauro" events.

It is thus interesting
to discuss the physical conditions for these phenomena to occur. Considering
the example of the Gaussian density matrix, it is seen from Fig.1  that the
behaviour
of the system is   mainly determined by the parameter $R\Delta$ and that
spectacular effects in charge ratios occur when $R\Delta$ is of order $1$
or smaller (as already noted, c.f. (\ref{21*}), $R\Delta\geq\frac12$). The
condition
\begin{equation}
R\Delta  \sim 1       \label{47}
\end{equation}
implies, for large multiplicity, either a very large particle and energy
density
(when $R$ is small and $\Delta$ takes its "canonical" value of $\sim
1fm^{-1}$),
or a "canonical" energy density of about $ 1$ Gev/fm$^3$ and a very small
average
momentum of the particles in the c.m. of the system. To be more specific, a
system of 100 pions satisfying (\ref{47}) at $\Delta =200$ MeV would
correspond to the energy density of about $35$GeV/4fm$^3 \approx 10$GeV/fm$^3$.
On the other hand, for an energy density of about $1$ GeV/fm$^3$ (and thus $R$
correspondingly
larger), $\Delta$ should not exceed 100 MeV.

Clearly, the probability of creating a "centauro" is enhanced if both effects
cooperate. We thus conclude that (a) the probabilty of creating a "centauro" is
enhanced in an environment of high energy density
 and that
the pions emerging from the "centauros" are likely to exibit abnormally small
relative momenta.

Finally, let us stress the crucial role played in this analysis by the
"original" Poisson multiplicity distribution $P_0(n)$ and by the reference
density
matrix $\rho_0(q,q')$ for the distinguishable particles. This distribution is
easily identified when working with Feynman diagrams\footnote{We thank J.Pisut
for a discussion about this point.}, but it is undefined experimentally. In
order to extract the HBT effect from experiment, one would like to compare the
data with a reference distribution where the HBT effect has been switched off.
The problem, how to define operationally this reference distribution has been
discussed for many years without a generally accepted conclusion
\cite{gg1,ha2}, for a review c.f. \cite{we1}. Consequently, it is not possible
to check separately the
assumptions about $\rho_0(q,q')$, $P_0(n)$ and the analysis of the HBT effect.
In the present paper we have chosen simple, but rather general, assumptions
about $\rho_0$, $P_0$ and concentrated on the HBT effect. No doubt, however,
more work remains to be done in order to find a realistic "distribution of
distinguishable particles" $\rho_0(q,q'), P_0(n)$.

\vspace{0.3cm}
{\bf Acknowledgements}
\vspace{0.3cm}

We would like to thank P.Bialas, J.Pisut and A.Staruszkiewicz for very useful
comments and discussions. This work was supported in part by the KBN Grant 2
P03B 08614.

\vspace{1.0cm}
{\bf Figure caption}
\vspace{0.3cm}

Fig.1. Frequency of "centauro" events as function of the average
total multiplicity for various values of the parameter $R\Delta$.

\end{document}